\documentclass[12pt,preprint]{aastex}
\usepackage{amsmath}
\usepackage{longtable}
\usepackage{subfigure}

\shorttitle{Polarized GRB Absorption Lines}

\shortauthors{Mao et al.}

\begin{document}


\title{On the Polarized Absorption Lines in Gamma-ray Burst Optical Afterglows}


\author{
J. Mao\altaffilmark{1,2,3}, R. J. Britto\altaffilmark{4}, D. A. H. Buckley\altaffilmark{5}, S. Covino\altaffilmark{6}, P. D'Avanzo\altaffilmark{6},
N. P. M. Kuin\altaffilmark{7}
}
\altaffiltext{1}{Yunnan Observatories, Chinese Academy of Sciences, 650011 Kunming, Yunnan Province, China}
\altaffiltext{2}{Center for Astronomical Mega-Science, Chinese Academy of Sciences, 20A Datun Road, Chaoyang District, Beijing, 100012, China}
\altaffiltext{3}{Key Laboratory for the Structure and Evolution of Celestial Objects, Chinese Academy of Sciences, 650011 Kunming, China}
\altaffiltext{4}{Department of Physics, University of the Free State, P.O. Box 339, Bloemfontein, 9300, South Africa}
\altaffiltext{5}{South African Astronomical Observatory, P.O. Box 9, Observatory Road, Observatory 7935, Cape Town, South Africa}
\altaffiltext{6}{Brera Astronomical Observatory, via Bianchi 46, I-23807, Merate (LC), Italy}
\altaffiltext{7}{Mullard Space Science Laboratory, Department of Space and Climate Sciences, University College London, Holmbury St. Mary, Dorking, RH5 6NT, UK}

\email{jirongmao@mail.ynao.ac.cn}

\begin{abstract}
Spectropolarimetric measurements of gamma-ray burst (GRB) optical afterglows contain polarization information for both continuum and absorption lines. Based on the Zeeman effect, an absorption line in a strong magnetic field is polarized and split into a triplet. In this paper, we solve the polarization radiative transfer equations of the absorption lines, and obtain the degree of linear polarization of the absorption lines as a function of the optical depth. In order to effectively measure the degree of linear polarization for the absorption lines, a magnetic field strength of at least $10^3$ G is required. The metal elements that produce the polarized absorption lines should be sufficiently abundant and have large oscillation strengths or Einstein absorption coefficients. We encourage both polarization measurements and high-dispersion observations of the absorption lines in order to detect the triplet structure in early GRB optical afterglows.
\end{abstract}




\section{Introduction}
The radiation mechanism is one of the important issues in gamma-ray burst (GRB) research. Synchrotron radiation is widely invoked to explain the nature of GRB afterglows \citep{piran04}. However, stochastic magnetic fields can be generated by the Weibel instability, and the radiation of relativistic electrons in the stochastic magnetic fields has different responses to the synchrotron radiation \citep{me99}. The hydrodynamical evolution of the radiative blast wave is strongly affected by the presence of magnetic fields (e.g., Sari et al. 1998; Mao \& Wang 2001a,b; Nava et al. 2013). Magnetized GRB outflows have been proposed in recent years, where GRB emission can be produced by magnetic dissipation \citep{lyutikov03,mc12,be14}. In the above scenarios, magnetic fields play a vital role in GRB physics.

Polarization can be a powerful tool to quantitatively analyze the strength and structure of the GRB magnetic fields.
In the high-energy band, GRBs observed by RHESSI, INTEGRAL, and IKALOS at early times have high degrees of linear polarization
\citep{co03,kalemci07,mcglynn07,yo11,yo12}, while POLAR recently detected some GRBs that show low degrees of linear polarization
\citep{zhang19,kole20}. In the optical band, polarimetric observations of GRB afterglows have been made \citep{covino99,wijers99,bersier03,steele09,uehara12,king14,gor16}. Time variability of linear polarization in GRB optical afterglows has been observed \citep{rol00,rol03,greiner03,wier12,mundell13,covino16,jm20}. Circular polarization in GRB optical afterglows has also been detected \citep{wier14}.

The observers studying GRB optical polarization usually focus on the GRB broadband polarimetry. It is important to note that spectropolarimetric measurements of optical GRBs provide simultaneous information on both continuum and absorption lines to further understand the physics of the GRB central engine and its environment. There has been recent progress in this interesting task. For example, the afterglow of GRB 191221B was successfully observed by multiwavelength observations. In particular, the spectropolarimetric measurements performed on both VLT/FORS2 and SALT are presented by \citet{buckley20}. A polarization degree of around 1\% is obtained for the continuum. Furthermore, the MgII (2796\AA) absorption line is clearly identified in the optical spectra. Unfortunately, the polarization is not well constrained for these weak absorption lines.

Many theoretical investigations on the polarization of GRB afterglows have been presented. For example, \citet{lan16} illustrated the polarization evolution in the early GRB afterglow. \cite{nava16} studied the circular polarization of the GRB afterglow in the optical band. \citet{mao18} performed the radiative transfer of the synchrotron polarization, and the model was applied to study the GRB afterglow polarization. The polarization radiative transfer of the GRB optical afterglow for the relativistic electron radiation in the random and small scale magnetic fields was presented by \citet{mao17}. But all the models mentioned above are focused on the GRB broadband polarimetry. In order to analyze the spectropolarimetric data of the GRB optical afterglows, we need polarization models for the GRB absorption lines.

Some absorption lines in GRB optical afterglows originate in the GRB natal regions, and they are produced by the highly ionized metals \citep{vree07,prochaska08,delia09,decia12,de12,heintz18}. Generally, three conditions are required for the production of the absorption lines. First, the continuum emission should be optically thin. Second, local thermodynamic equilibrium should be satisfied during the process of the absorption line production. Third, the elements should have large number densities and transitions with a large number of weighted oscillation strengths or Einstein absorption coefficients, so that significant line features can be identified with a good signal-to-noise (S/N) ratio. In particular, we discuss two cases where the absorption lines can be affected by magnetic fields in GRB afterglows. First, GRB forward shock can sweep up the GRB surrounding medium and produce a continuum in the ultraviolet (UV) and optical bands \citep{piran04}. Due to photoionization, absorption lines are produced in the natal regions of GRBs. Then, magnetic fields can be strongly amplified by the GRB shock front in a relatively small area. The absorption lines in the small area are affected by the amplified magnetic field. Second, GRB outflow can be strongly magnetized \citep{lyutikov03}. When the magnetically dominated GRB outflow encounters the dense medium that is located ahead of the outflow, the magnetized outflow may provide the UV and optical continuum \citep{be14}. Some absorption lines can be produced due to photoionization. Then, the absorption lines are affected by the magnetic field of the GRB outflow. In both cases mentioned above, a single absorption line in contrast to the continuum in a strong magnetic field can be transformed into a triplet. This is the so-called Zeeman effect.

The split in the absorption lines is due to polarization. The polarization radiative transfer of the Zeeman triplet was first presented by \citet{unno56}.
Some applications have been shown in the astrophysical field. Stellar magnetic fields can be measured by the multiplet lines \citep{takeda91}. Lines enhanced by the Zeeman effect were detected in some extremely active K dwarfs \citep{basri94}. The linear and circular hydrogen lines in the white dwarfs were carefully analyzed \citep{donati94,ach95}, and the CaI and CaII K and H lines were also detected \citep{kawka14}. The magnetic field structure of white dwarfs can be obtained by the Zeeman effect reconstruction \citep{euchner02}. In this paper, we use the Zeeman effect of the absorption lines in the GRB optical afterglow. This is the first time that the Zeeman effect in the GRB research has been proposed.

We present the polarization radiative transfer process of the absorption lines due to the Zeeman effect in Section 2.1. The results in different absorption cases with different physical parameters are given in section 2.2. In section 3, we discuss the GRB magnetic fields and the conditions of the Zeeman triplet in GRB afterglows. The observational suggestions to identify the polarized metal elements are also presented. A brief conclusion is given in Section 4.

\section{Radiative Transfer of Absorption Line by Zeeman Effect}
\subsection{Formalism}
Polarized radiation is produced by electron oscillation in magnetic fields. Because of the Zeeman effect, the oscillators are denoted by the $p$ (along the magnetic field direction), $l$ (left-rotated), and $r$ (right-rotated) electrons. Therefore, unpolarized radiation becomes polarized after the transfer through the magnetic field. We adopt the formulation by \citet{unno56}. The equations of the radiative transfer for the Stokes parameters of $I$, $Q$, and $V$ are given as:
\begin{equation}
\cos\theta\frac{dI}{d\tau}=(1+\eta_I)I+\eta_qQ+\eta_vV-(1-\eta_I)E,
\end{equation}
\begin{equation}
\cos\theta\frac{dQ}{d\tau}=\eta_QI+(1+\eta_I)Q-\eta_QE,
\end{equation}
and
\begin{equation}
\cos\theta\frac{dV}{d\tau}=\eta_VI+(1+\eta_I)V-\eta_VE.
\end{equation}
Here, the stokes parameter $U$ is set to be zero, and we obtain the degree of linear polarization to be $Q/I$. We assume that the initial stage is unpolarized. $E$ is the emission in the medium presented in the above equations. Here, we focus on the absorption effect by the medium, and we simply neglect the emission effect. Then we set $E=0$.
$\tau$ is the optical depth of the continuum, and $d\tau=k\sec\theta dz$, where $k$ is the absorption coefficient of the continuum, $z$ is the direction of the light propagation, and $\theta$ is the angle between the light propagation direction and the normal direction of the absorption plane.
$\eta_I$, $\eta_Q$, and $\eta_V$ are the parameters related to the absorption line coefficient and the continuum absorption coefficient. They are calculated by
\begin{equation}
\eta_I=\frac{\eta_p}{2}\sin^2\phi+\frac{\eta_l+\eta_r}{4}(1+cos^2\phi),
\end{equation}
\begin{equation}
\eta_Q=(\frac{\eta_p}{2}-\frac{\eta_l+\eta_r}{4})\sin^2\phi,
\end{equation}
and
\begin{equation}
\eta_V=(\frac{-\eta_l+\eta_r}{2})\cos\phi,
\end{equation}
where $\phi$ is the angle between the direction of the magnetic field and the direction of the light propagation. We define $\eta_p=k_p/k$, $\eta_l=k_l/k$, and $\eta_r=k_r/k$, where $k_p$, $k_l$, and $k_r$ are the absorption coefficients of the $p$, $l$, and $r$ electrons, respectively. The absorption coefficients are dependent on the wavelength. We further assume a Gaussian form as $\eta_p=\eta_0\exp(-u^2)$, $\eta_l=\eta_0\exp[-(u-u_B)^2]$, and $\eta_r=\eta_0\exp[-(u+u_B)^2]$, where $\eta_0$ is the ratio between the line absorption coefficient and the continuum absorption coefficient, $u=(\lambda-\lambda_0)/\Delta\lambda_D$, and $\Delta\lambda_D$ is the line width. We simplify the case in which the polarization is in the line center, and we have $u=0$. $u_B$ is defined as $u_B=\Delta\lambda_B/\Delta\lambda_D$, and $\Delta\lambda_B=eB\lambda^2/4\pi m_0c^2$ is the wavelength shift because of the Zeeman effect. In the calculation, we assume that $\phi$, $\theta$, $\eta_0$ and $u_B$ are free parameters to be adjusted. Because GRB afterglow occurs in a region in which GRB shocks sweep up the medium in a short dynamical timescale, and the afterglow emission region is optically thin,
we only solve the equations (1), (2), and (3) once, and the Milne-Eddington method is not necessary.

The Zeeman effect provides the wavelength shift of $\Delta\lambda_B=eB\lambda^2/4\pi m_0c^2$, and we have $u_B=\Delta\lambda_B/\Delta\lambda_D=eB\lambda^2/4\pi m_0c^2\Delta\lambda_D$. The spectral resolution is $R=\lambda/\Delta\lambda_D$, then we obtain $u_B=eBR\lambda/4\pi m_0c^2$. Here, $\lambda$ is the observational wavelength. The observational wavelength is redshifted as $\lambda=\lambda_0(1+z)$, where $\lambda_0$ is the intrinsic wavelength. Therefore, we can constrain the magnetic field as
\begin{equation}
B=2.6\times10^3(\frac{u_B}{1.0})(\frac{R}{1.0\times 10^5})^{-1}(\frac{\lambda_0}{4089\AA})^{-1}(\frac{1+z}{2.0})^{-1}~\rm{G}.
\end{equation}
Here, we take the SiIV(4089\AA) absorption line at redshift $z=1.0$ as an example, and $u_B$ is assumed to be 1.0. Because the spectral resolution is $10^5$ in the above equation, we note that the Zeeman effect can be effectively detected by high-dispersion spectrographs.

\subsection{Results}
We solve the Equations (1), (2), and (3) and obtain the results of the radiative transfer for the absorption lines.
The degree of linear polarization as a function of the optical depth is shown in Figures 1 and 2. In general, we see that the polarization degree is linearly increased with the optical depth of the medium.
We specify the following results.
\begin{itemize}
\item[-]The polarization degree increases when the angle $\phi$ increases. This indicates that the polarization degree is dependent on the magnetic field morphology and structure. The polarization is weakest when the magnetic field direction is parallel to the light propagation direction, and the polarization is strongest when the magnetic field direction is perpendicular to the the light propagation direction.
    \item[-]The polarization degree increases when the angle $\theta$ increases. This indicates that the photons of absorption lines passing through the absorption medium with longer distance suffer stronger magnetic field effects and have stronger polarization.
        \item[-]The polarization degree increases when $\eta_0$ increases. This means that the stronger polarization of the absorption lines can be obtained when the absorption lines are stronger.
            \item[-]The polarization degree increases when $u_B$ increases. Obviously, the polarization of the absorption lines is stronger when the magnetic field is stronger. In the results mentioned above, we see that the polarization degree increases when the optical depth of the continuum increases. When the optical photons suffer stronger absorption in the medium, the magnetic field effect on the absorption lines is stronger and the polarization of the absorption lines is stronger.
\end{itemize}

When we compare the specified results mentioned above, it seems that the angle $\phi$ has a strong effect on the polarization degree that is evolved with the optical depth of the medium. This means that the magnetic field morphology is very important for the radiative transfer of the GRB absorption line polarization. Compared to $\phi$, the other parameters, such as $\theta$, $\eta$, and $u_B$, are less important for the polarization degree of the GRB absorption lines. It makes us further consider the magnetic field structure in the GRB outflows.

We encourage spectropolarimetric measurements to GRB afterglows in the optical band. We expect that the polarization degree of the absorption lines can be obtained. We can also attempt to detect the triplet structure in the absorption lines, and a telescope with a spectral resolution of about $10^5$ is necessary to detect the structure with good S/N ratio.
The polarization of the absorption lines can reach an average degree of a few percent. In this case, a magnetic field strength of about $10^3$ G is required. In addition, although the polarization degree is larger if the optical depth of the continuum is larger, the detection of the absorption line in GRB optical afterglows prefers a case with small optical depth. Thus, a moderate absorption is suitable for the detection of the polarized absorption line (see the detailed discussion in Section 3).

\section{Discussion}

Strong polarization of absorption lines can be achieved when the optical depth of the continuum is large. However, GRB afterglows are too faint to detect if the optical depth of the continuum is very large. At present, we do not have any systematic analysis of the optical depth for GRB afterglows. Here, we do a simple estimation. The optical depth is defined as $\tau=k_\nu nl$, where $n$ is the number density, $k_\nu$ is the atom absorption coefficient for the continuum, and $l$ is the length that photons pass through. In principle, we should sum up the contribution from all of the atoms to calculate $k_\nu$. For simplicity, it is reasonable to take $k_\nu=6.3\times 10^{-18}~\rm{cm^2}$, which is obtained from only the neutral hydrogen photoionization at the energy level of the ground state. We ignore the absorption effects due to the photoionization by other atoms. The absorption due to the scattering effect is also neglected. We take the length $l$ as the thickness of the GRB surrounding medium. Because the GRB surrounding medium is swept by GRB shocks, the length can be estimated by $l=r/\Gamma^2$, where $r$ is the fireball radius, and $\Gamma$ is the bulk Lorentz factor. We obtain
\begin{equation}
\tau=k_\nu nl=0.3(\frac{n}{1.0\times 10^5~\rm{cm^{-3}}})(\frac{r}{1.0\times 10^{15}~\rm{cm}})(\frac{\Gamma}{50.0})^{-2}.
\end{equation}
It indicates that the optical depth should be moderate to satisfy the detection for both continuum and polarization of absorption lines.

We see in Section 2.2 that the polarization radiation transfer of the absorption lines in GRB optical afterglows is strongly related to the magnetic field structure. The polarization
can be strongest when the magnetic field direction is perpendicular to the light propagation direction. The magnetic field structure in the GRB surrounding medium is quite uncertain. When we consider a large scale magnetic field in a magnetized GRB outflow, the magnetic field can be decomposed as toroidal and poloidal components. The toroidal component can affect on the polarization radiation transfer of the absorption lines.

We can attempt to constrain the magnetic field strength by the polarization and Zeeman splitting of the absorption lines in GRB optical afterglows. \citet{mizuno14} performed some numerical simulations to identify the amplified magnetic field behind the relativistic shocks. In such cases, the absorption lines are strongly affected by the magnetic fields. The magnetic field in the GRB afterglows can be simply estimated as \citep{sari98}:
\begin{equation}
B=4.3\times 10^3(\frac{\epsilon_B}{0.5})^{1/2}(\frac{n}{1.0\times 10^5~\rm{cm^{-3}}})^{1/2}(\frac{\Gamma}{50.0})~\rm{G},
\end{equation}
where $\epsilon_B$ is the ratio between the magnetic energy density to the total energy density behind the GRB shock, $n$ is the number density, and $\Gamma$ is the bulk Lorentz factor. \citet{prochaska08} estimated that the number density should be at least larger than $10^3~\rm{cm^{-3}}$ in the photoionization scenario. The stellar wind environment provides a large number density in GRB afterglows. Similarly the clumpy medium surrounding the GRB progenitor can also be very dense with a large number density, and a reverse shock is often generated in this case. Therefore, strong magnetic fields in GRB afterglows can be achieved.
On the other hand, when GRB outflows are magnetized, the magnetic field in the reverse shock region is stronger than that in the forward shock region \citep{fraija15}. If we consider the magnetized outflow, the magnetization parameter can be defined as $\sigma=B^2/\gamma nm_ec^2$. Here, $\gamma$ is the Lorentz factor of the electrons, and $m_e$ is the electron mass. We take $\sigma=1.0\times 10^3$ as a reference value. The magnetic field can be estimated as
\begin{equation}
B=8.7\times10^3(\frac{\gamma}{1.0\times 10^6})^{1/2}(\frac{n}{1.0\times 10^5\rm{cm^{-3}}})^{1/2}(\frac{\sigma}{1.0\times 10^3})^{1/2}~\rm{G}.
\end{equation}
This presentation is not a strong constraint to the magnetic field strength. It only means that GRB outflow can be highly magnetized, even when the plasmas are dense and relativistic. In such cases, the magnetic field strength can affect on the polarization of the absorption lines.
The magnetic field number estimated by Equations (9) and (10) is roughly consistent with that in Equation (7). The estimation of the magnetic field strength is related to some parameters, such as the bulk Lorentz factor, the particle number density, and the total energy density in the shocks. These parameters cannot be constrained by a single polarimetric measurement. For each GRB, it is necessary to perform multiwavelength observations to constrain these parameters and try to break the possible degeneracies among these parameters. If we take the numbers of the magnetic field strength estimated by Equations (9) and (10), we predict that the Zeeman effect with this magnetic field strength level has a possibility of being detected by large optical telescopes \citep{bagnulo18}.
However, we note that magnetic fields can be sustained only for a short time before they quickly decay \citep{rossi03}. Magnetic dissipation is important when GRB outflow is magnetized, and synchrotron cooling seems too efficient \citep{be14}. It seems that the polarized absorption lines with the significant Zeeman effect structure can only appear over a short timescale.

In order to further present the polarization degree that is related to the magnetic field strength, we show the result in Figure 3. The case has the conditions of $\phi=30^\circ$, $\theta=30^\circ$, and $\eta_0=1.0$. The polarization degree increases when the magnetic field strength increases from 100 to about $10^4$ G. When the magnetic field is larger than about $10^4$ G, $\eta_r$ and $\eta_l$ approach to be zero. This means that the further polarization by the left- and right-rotated electron oscillation is stopped, and the polarization degree remains constant. In the meanwhile,
the magnetic fields calculated by Equations (9) and (10) are also marked in Figure 3. This indicates that the polarization degree of the GRB absorption lines can reach a level of about 6-7\% degree in normal GRB afterglow cases.

Although we focus on the research of the polarization for some absorption lines in GRB afterglows, the mechanisms to generate the highly ionized absorption lines should be discussed. In this paper, we assume that the absorption lines originated from the GRB afterglow photoionization \citep{perna98}. When the GRB fireball produces forward shocks, the shocks sweep up the surrounding medium. \citet{delia09} measured the FeII absorption line in the GRB 080330 afterglow\footnote{Although the topic of this paper is the polarized absorption lines in the optical band, the X-ray absorption can provide some hints about the
metal abundance. We take this GRB as an example. It was triggered by Swift and the observation was performed by the X-ray telescope in time (Mao et al. 2008a,b). The intrinsic column density of $N_{\rm{H,X}}=5.1^{+3.1}_{-2.8}\times 10^{21}{\rm{cm^{-2}}}$ is determined by the early X-ray spectrum. This large number indicates the strong absorption by the metals nearby the burst.}, and the absorption region was located at about a few kiloparsecs from the burst center. \citet{prochaska08} detected some NV absorption lines in a spectral sample of GRB optical afterglows. The photoionization region is determined to be about 10 pc from the burst center, and the GRB environment has a number density larger than $10^3~\rm{cm^{-3}}$.
\citet{prochaska08} also suggested that some other elements, such as CIV, SiIV, and OIV, could be detected due to the GRB afterglow photoionization. \citet{decia12} performed the rapid-response observation to GRB 080310 by VLT/UVES. The time variability of the FeII absorption lines was clearly identified. This confirms the process of the gradual photoionization by GRB optical afterglows. However, the time variability of the NV absorption line was not detected. It seems that the photoionization process is quite complex.
On the other hand, if the GRB outflow is magnetized, the outflow may encounter the clumpy material in the direction of the outflow propagation. Thus, the photoionization region in which the absorption lines can be produced is not limited to be at 10 pc, and more highly ionized elements can be identified in GRB optical afterglows. We note that the located region mentioned above might not be the region in which magnetic fields are amplified. The production of the absorption lines in GRB afterglows depends on the metal ionization potentials. We investigate the ionization potentials for some elements. They are CIV(64.5eV), OII(35.1eV), OIII(54.9eV), OIV(77.4eV), SiIII(33.5eV), SiIV(45.1eV), and NV(97.9eV). We note that
the highly ionized ions, such as CIV, OIII, OIV, and NV, are generated by photoionization from GRB progenitors or shocks, while shock heating is neglected in GRB afterglows. In addition, we assume that metal elements are inside the shocked region where magnetic fields are amplified. However, some elements producing highly ionized absorption lines may not in the region of the amplified magnetic field.
Moreover, photoionization is not a unique mechanism to produce GRB absorption lines. \citet{heintz18} performed a comprehensive investigation on highly ionized absorption lines, and using recombination processes to produce absorption lines was
proposed. However, the recombination timescale is estimated
to be about 1 day, and it is larger than the photoionization timescale \citep{perna02}.
Therefore, to further investigate the location of photonionization or recombination processes for a certain element is an important issue for future research of the GRB polarized absorption lines.

There should be enough abundance for a certain element in the dense medium or the GRB ejecta, such that the element can be identified in a spectrum. However, GRB nucleosynthesis related to hydrodynamics and microphysics in the GRB central engine is quite complicated. For example, \citet{be03} initially investigated the nucleosynthesis conditions in the GRB fireballs. \citet{janiuk14} studied the GRB nucleosynthesis processes, and some metal elements, such as Si, S, Cl, Ar, Ca, Ti, and Fe, with their isotopes, were presented. Here, we neglect the rapid electron-capture process ($r$-process) elements because they have very small abundances. In principle, some metals from the GRB nucleosynthesis can be shown in the absorption spectra of the GRB afterglows \citep{perna98,heintz18}. In addition, supernova explosion may enrich the GRB surrounding medium with heavy elements \citep{decia12}. The $\alpha$-process elements with their isotopes in the supernova explosion nucleosynthesis can be referenced \citep{woosley73}.
Furthermore, oscillation strength is the other important factor for the detection of absorption lines. A larger oscillation strength number of one element indicates a stronger absorption. For the forbidden lines, a larger number of the Einstein coefficient corresponds to a stronger absorption. Therefore, one element with both large abundance and large oscillation strength/Einstein absorption coefficient number is expected to produce a strong polarized absorption line in strong magnetic fields. We
list our suggestions of the absorption lines in Tables 1 and 2\footnote{Although some absorption lines may not be produced by highly ionized elements, we still list them because they are often shown in GRB optical spectra.}.
From the observational point of view, a large sample of GRB optical spectra has been obtained, and the equivalent widths (EWs) of the absorption lines in the spectra have been statistically analyzed \citep{de12}. The species, such as SiII(1527), SiIV(1394, 1403), CIV(1549), FeII(1608, 2374, 2587), AlII(1671), AlIII(1855, 1863), MgII(2800), and CaII(3935, 3970), were clearly recognized with the effective EWs. In particular, CIV and SiIV have the rest-frame EWs of about 1 \AA. Furthermore, some highly ionized metals inside the natal regions of GRBs were identified in some optical spectra, and the elements, such as NV, OVI, CIV, and SiIV, were noticed by \citet{heintz18}. However, we note that some elements have small oscillation strength numbers: NV(1239), NV(1243), CIV(1548), CIV(1551), CIV(5801), OVI(1032), and OVI(1038) have the weighted oscillation strength ($log(gf)$) values of $-0.505$, $-0.807$, $-0.419$, $-0.721$, $-0.194$, $-0.576$, and $-0.879$, respectively.

Based on the analysis mentioned above, we have a few expectations for the detection of the GRB polarized absorption lines. First, luminous UV-optical afterglows are required to photoionize and produce significant absorption lines. Second, a spectropolarimetric observation should have moderate optical depth in order to detect both continuum and absorption line polarization.
Third, a sufficiently strong magnetic field (larger than $10^3$ G) is presented at the onset of the GRB afterglows and a successful spectropolarimetric observation should be performed within a certain time before the magnetic field has decayed. One may perform the spectropolarimetric observations at different epochs to measure the time variability of the polarized absorption lines.
Fourth, the metal elements may have large number densities and large weighted oscillation strengths/Einstein absorption coefficients. The sufficiently strong absorption lines (i.e. the rest-frame EWs are larger than 1 \AA) should be presented in the spectra.

GRB optical afterglows usually show modest polarization \citep{covino16}.
Although observers were likely unaware of the detection possibility of polarized absorption lines in GRB afterglows, a thorough examination of observational samples could be opportune.
Some spectral techniques have been further developed. For example, \citet{sen10} presented the Zeeman component decomposition method to extract the Zeeman component profile in the polarized spectra. All Stokes parameters can be simultaneously obtained by solving the Zeeman component profile, and the line-of-sight magnetic field can be also constrained. This method was successfully applied to measure the magnetic field with the strength of less than 1 G in a very bright K giant \citep{sen11}. Although the detection of line polarization and Zeeman splitting in GRB optical afterglows is challenging for spectropolarimetric observations, the detection with some special data analysis methods would provide important results. 
On the other hand, because the
spectral resolution is given to be $10^5$ in Equation (7), we strongly suggest that the telescopes with the high-dispersion spectrograph can be applied to obtain the high-resolution spectra of the GRB optical afterglows. It is worth attempting to detect the Zeeman splitting structure by the high-dispersion observations.

The polarization of the continuum and the polarization of absorption lines can originate in different emission regions, and they are explained by different radiation mechanisms. It is difficult to investigate different polarization origins by just broadband polarimetry, though the spectropolarimetric observation is suitable to examine the polarization processes on both continuum and absorption lines. Furthermore, GRB afterglows usually evolve rapidly with time, posing a challenge to the detection of polarization. Number densities of metal elements, magnetic field structure and strength, and emission geometry of GRB afterglows also evolve with time.
Thus, the calculation of the absorption line polarization in this paper is time-averaged. We encourage multiple spectropolarimetric observations in the future.

We focus on the polarization in the center of an absorption line. However, the polarization from the wing of an absorption line can be included when one performs polarization measurements. We list three minor effects on the polarization measurements of absorption lines below. First, we may detect the polarization generated by other bright optical sources nearby GRB sources. Second, some polarized photons are emitted from the GRB host galaxy. Third, some polarized photons are emitted from the interstellar medium (ISM) and the intergalactic medium (IGM).
However, we think that the effects mentioned above have minor contributions to the intrinsic polarization, and the polarization from these minor effects can be neglected.

\section{Conclusions}
We obtain the polarization of absorption lines by solving the polarization radiative transfer equations for GRB optical afterglows. A few percent of linear polarization degree can be detected when magnetic fields reach a value of about $10^3$ G. We suggest some highly ionized absorption lines, and the metal elements to produce these absorption lines should have enough abundances and large oscillator strengths/Einstein absorption coefficients for the spectropolarimetric observation. The detection of the Zeeman splitting structures in absorption lines by the high-dispersion spectrographs are expected. We encourage more spectropolarimetric investigations of the early GRB optical afterglows in the future.

\acknowledgments
We are grateful to the referee for a careful review and very helpful suggestions.
We thank Drs. K. Wiersema, C. G. Mundell, Z. Liu, Y.-H. Zhao, and X.-L. Yan for their insightful discussions.
J.M. is supported by the National Natural Science Foundation of China 11673062 and the Oversea Talent Program of Yunnan Province. D.A.H.B acknowledges the support through the National Research Foundation (NRF) of South Africa. N.P.M.K acknowledges the support by the UK Space Agency.

\clearpage




\clearpage

\begin{figure}
\center
\includegraphics[scale=0.4]{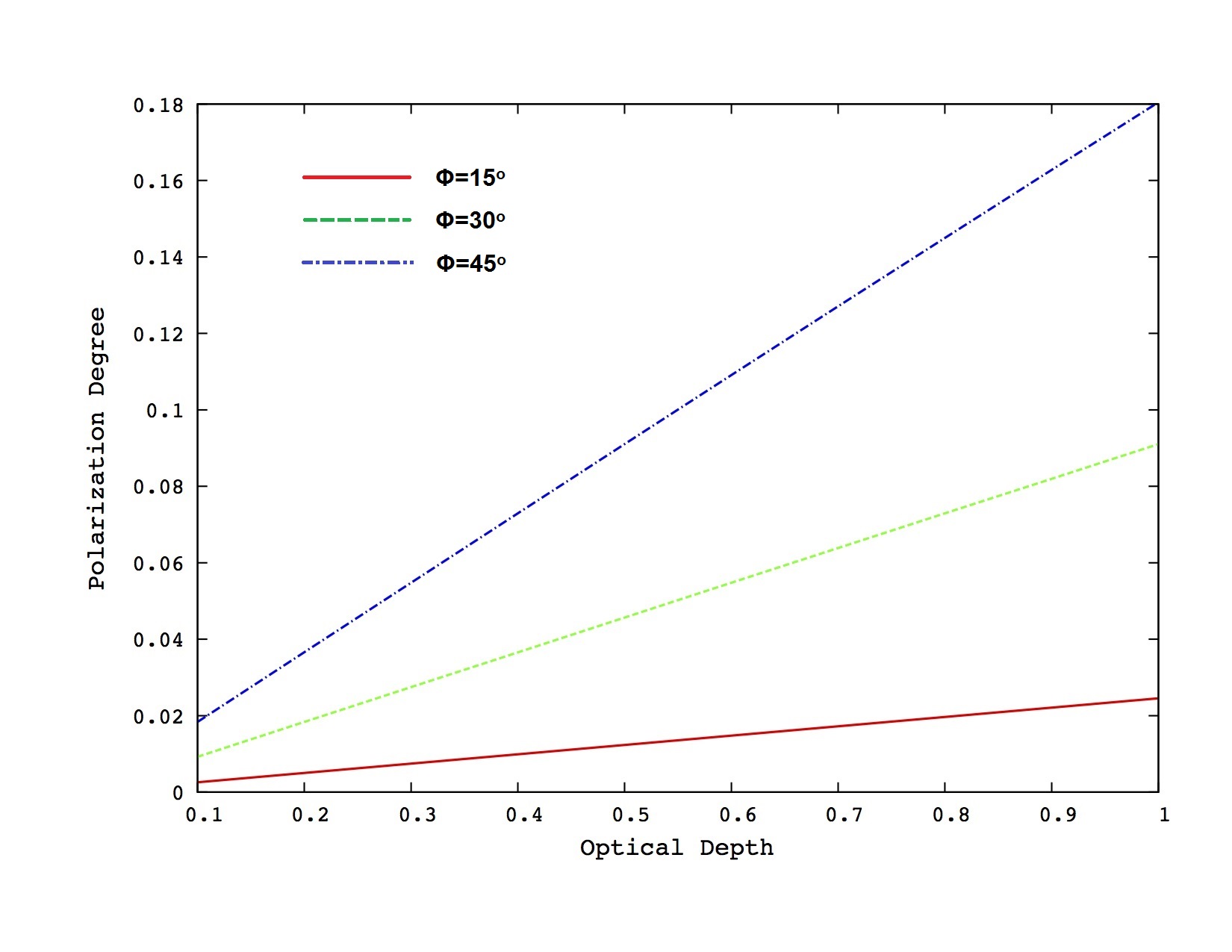}
\includegraphics[scale=0.4]{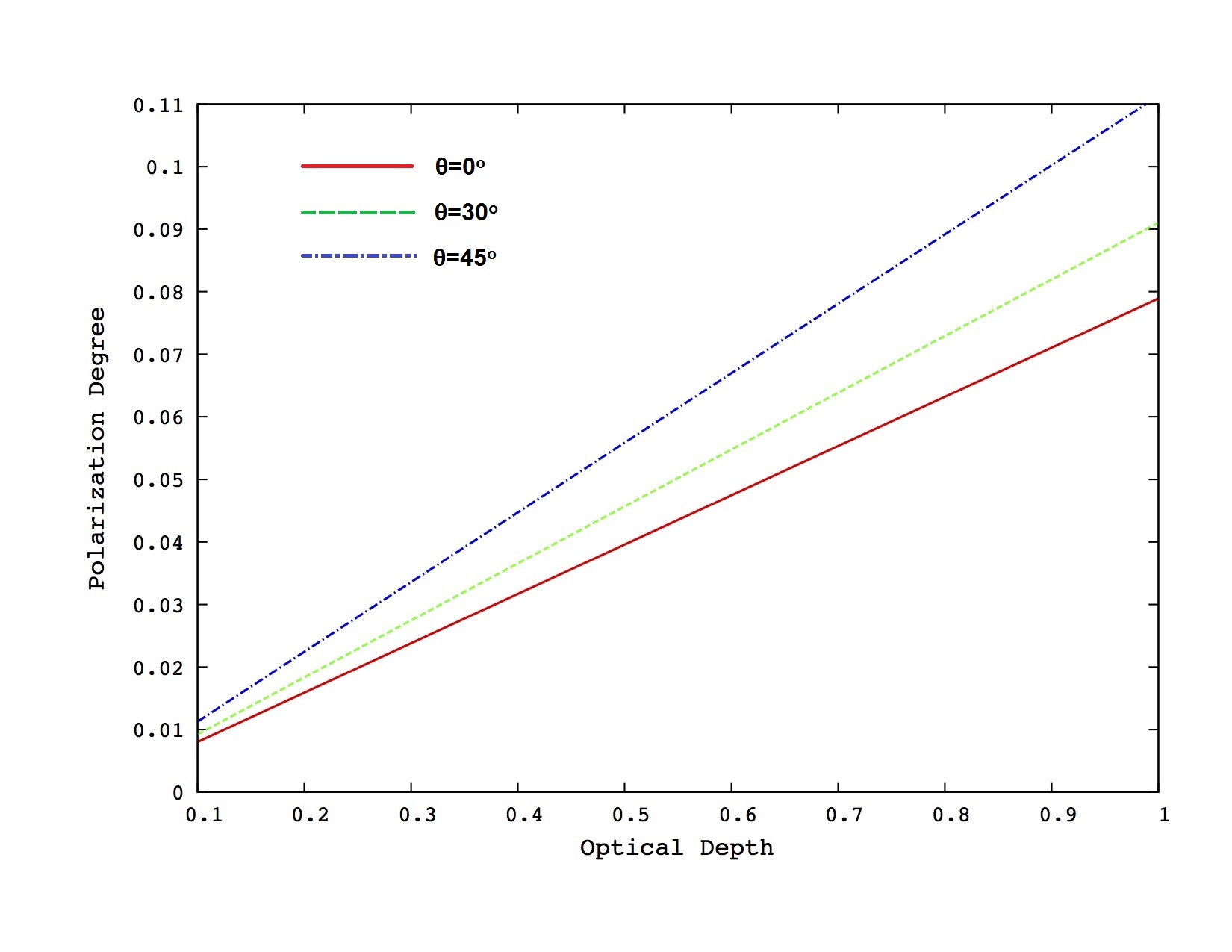}
\caption{The polarization degree as a function of the optical depth (I). Upper panel: the results are given by $\theta=30^\circ$, $u_B=1.0$, and $\eta_0=1.0$. The cases of $\phi=15^\circ,~30^\circ,~\rm{and}~45^\circ$ are indicated by solid (red), dashed (green), and dashed-dotted (blue) lines. Lower panel: the results are given by $\phi=30^\circ$, $u_B=1.0$, and $\eta_0=1.0$. The cases of $\theta=0^\circ,~30^\circ,~\rm{and}~45^\circ$ are indicated by solid (red), dashed (green), and dashed-dotted (blue) lines.
\label{fig1}}
\end{figure}

\begin{figure}
\center
\includegraphics[scale=0.4]{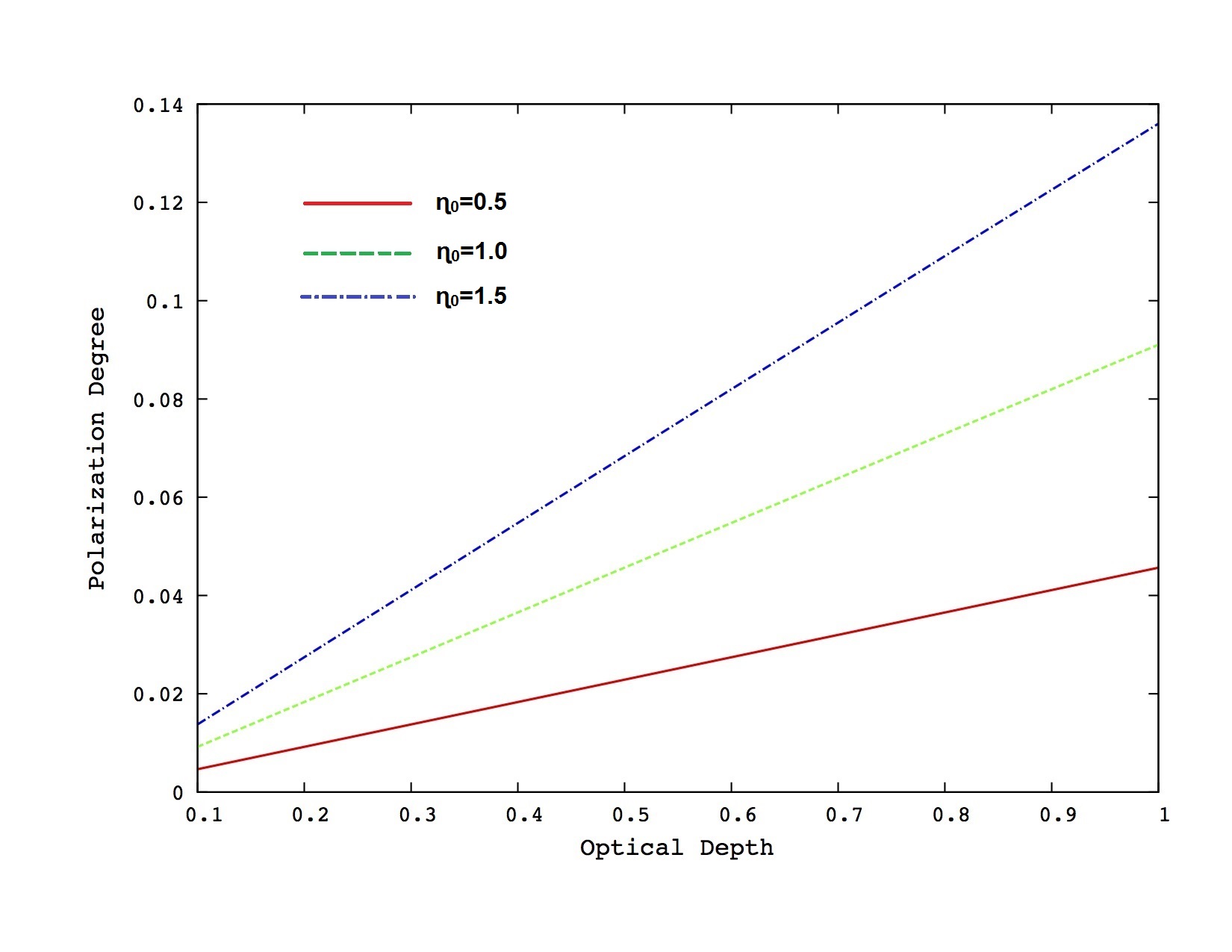}
\includegraphics[scale=0.4]{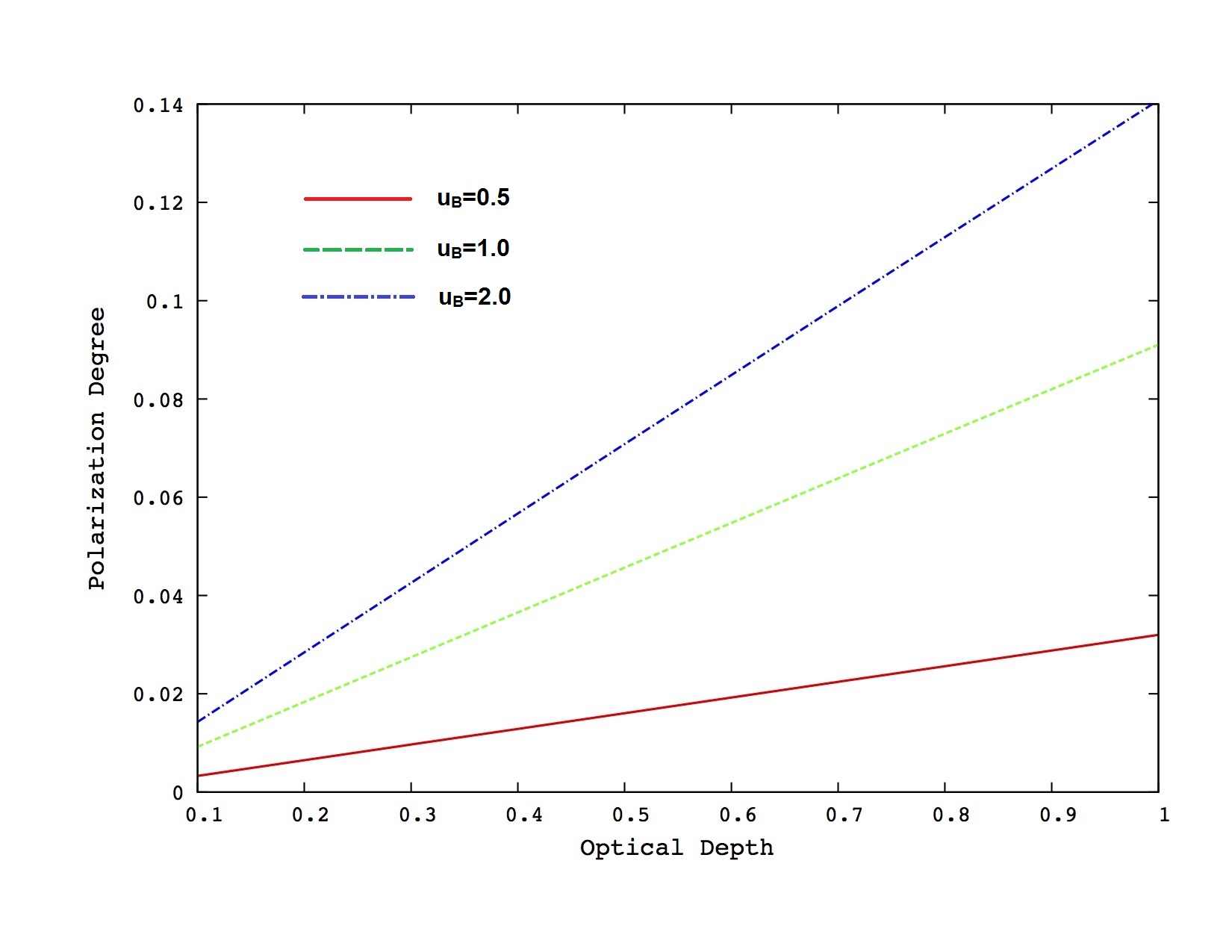}
\caption{The polarization degree as a function of the optical depth (II).
 Upper panel: the results are given by $\theta=30^\circ$, $\phi=30^\circ$, and $u_B=1.0$. The cases of $\eta_0=0.5, 1.0,~\rm{and}~1.5$ are indicated by solid (red), dashed (green), and dashed-dotted (blue) lines. Lower panel: the results are given by $\theta=30^\circ$, $\phi=30^\circ$, and $\eta_0=1.0$. The cases of $u_B=0.5, 1.0,~\rm{and}~2.0$ are indicated by solid (red), dashed (green), and dashed-dotted (blue) lines.
\label{fig1}}
\end{figure}

\begin{figure}
\center
\includegraphics[scale=0.5]{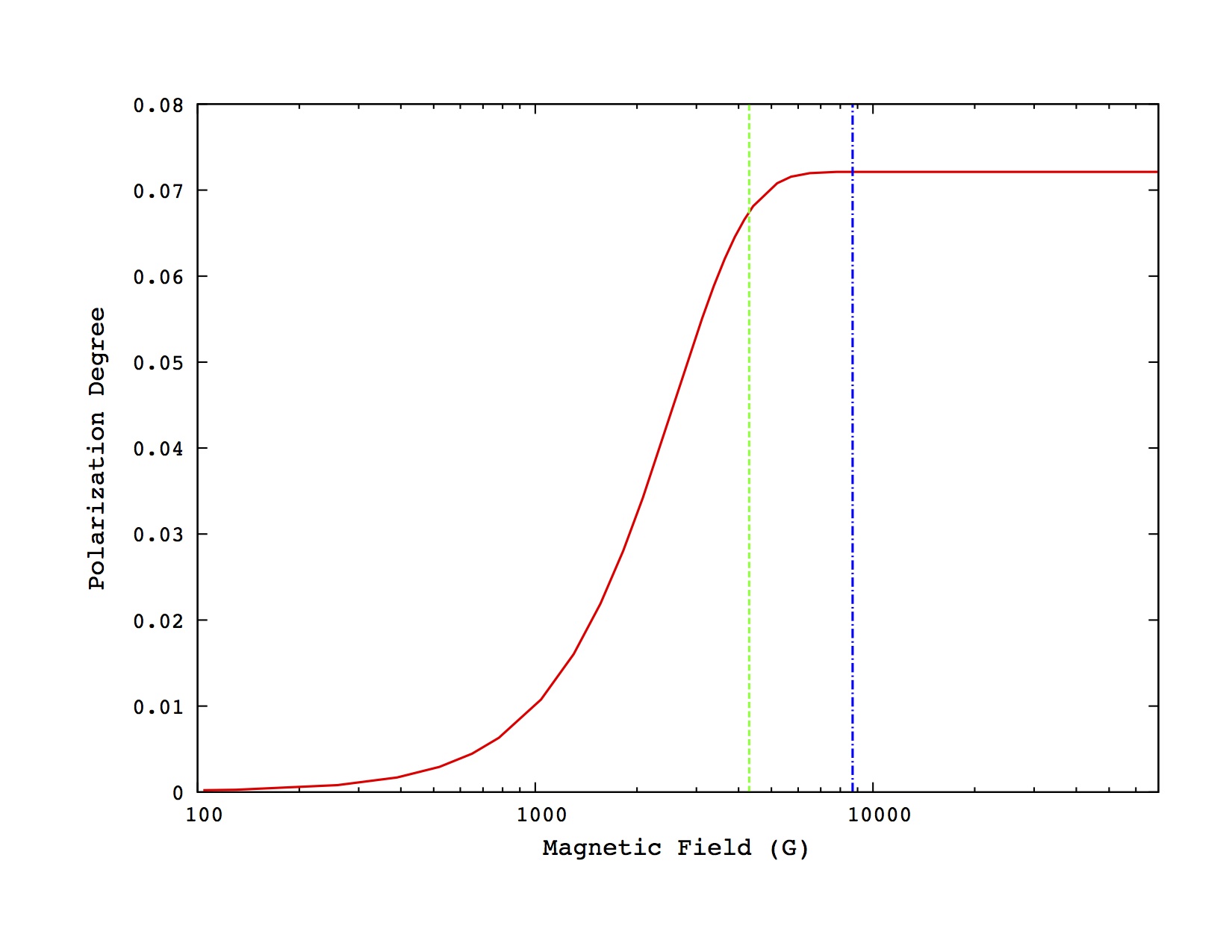}
\caption{The polarization degree as a function of the magnetic field. The solid line (red) presents the result that is given by $\phi=30^\circ$, $\theta=30^\circ$, and $\eta_0=1.0$. The magnetic fields calculated by Equations (9) and (10) are labeled by dashed (green) and dotted-dashed (blue) lines, respectively.
\label{fig1}}
\end{figure}

\clearpage

\begin{table}
\scriptsize{
\begin{tabular}{ccccccccccccccccc}
\hline
Line &Wavelength &IP& $log(gf)$ & Line & Wavelength &IP& $log(gf)$ & Line & Wavelength &IP& $log(gf)$ \\
     &\AA        & eV          &   &   & \AA        &    eV       &   &   &\AA         &  eV         \\
  \hline
CII  & 904.1 &11.3&0.224 & CII & 7236 &11.3& 0.298 & NII & 3995 &14.5 & 0.215 \\
NII  &5679.6 &14.5&0.250 & NII &500.2 &14.5& 0.592 & NIII& 4515 &29.6 & 0.211 \\
NIV  &3481   &47.4&0.238 & OII & 4649 &13.6& 0.307 & OII & 4119 &13.6 & 0.433  \\
OIII &3760   &35.1&0.162 & OIII& 3715 &35.1& 0.149 & OIV & 238.6&54.9 & 0.258 \\
OIV  &3386   &54.9&0.148 & NeVI& 122.7&126.2& 0.313& MgII& 4481 &7.6 & 0.973 \\
MgII & 2796  &7.6&0.09  & AlII& 1671 &6.0 &0.263 & AlIII& 5696&18.8 & 0.235 \\
SiII & 6347  &8.2&0.23  & SiII& 4131 &8.2 &0.463 & SiII & 1195& 8.2& 0.49 \\
SiII & 1265  &8.2&0.52  &SiIII& 1207 &16.3&0.22   & SiIII& 4553&16.3 & 0.292 \\
SiIV &4089   &33.5&0.195 &SV   & 786.5&47.2& 0.165 &SV    & 661.5&47.2& 0.802  \\
CaII & 3934  &6.1&0.135 &CaII & 3179 &6.1& 0.51  & TiII & 3361 &6.8& 0.28 \\
TiII & 3235  &6.8& 0.336\\
 \hline
\end{tabular}}
\caption{\footnotesize{The allowed lines suggested to have the Zeeman effect in strong magnetic field. IP is the ionization potential, and $log(gf)$ is the weighted oscillation strength. The parameters are taken from Allen's astrophysical quantities in the fourth edition.}}
\end{table}

\begin{table}
\scriptsize{
\begin{tabular}{cccccccccccccccc}
\hline
Line &Wavelength &IP& $A$ & Line & Wavelength &IP & A & Line & Wavelength &IP& A \\
     &\AA     & eV   &  s$^{-1}$  &      & \AA   & eV     &    s$^{-1}$       &      &\AA  &eV       &    s$^{-1}$       \\
  \hline
$[$NII]  & 5755 &14.5&1.17 & [OIII] & 4363 & 35.1& 1.71 & [NeIII] & 3342 &41.0& 2.72 \\
$[$NeIII]& 1794 &41.0&1.88 & [NeIII] & 1815&41.0 & 2.02 & [NeIV]  & 1612 &63.4& 1.23 \\
$[$NeV]  &2973  &97.1&2.89 & [SIII] & 6312 &23.3& 2.22 & [ArIII]  & 5192 &27.6& 2.59  \\
$[$ArIII] &3109 &27.6&3.91 & [ArIV]& 2854  &40.7& 2.11 & [ArV]    & 4626 &59.7& 3.29 \\
$[$ArV]  &2691  &59.7&6.55 & [CaV]& 5309&67.2& 1.90 & [CaXII] & 3328&108.8& 487 \\
$[$CaXIII]&4087 &127.2&319  & $[$FeII] & 4890  &7.9& 0.36 &[FeII]& 4416 &7.9& 0.46 \\
$[$FeII] & 4287 &7.9& 1.5 &$[$FeII] & 4359  &7.9 &1.1  &[FeII] & 5528 &7.9 &0.12   \\
$[$FeII]& 4244  &7.9& 0.90 & $[$FeIII] & 4658 &16.2&0.44 &[FeIII] & 5270&16.2& 0.40 \\
$[$FeIV]& 4907  &30.7& 0.32 &$[$FeV] & 3895  &54.8&0.71 &[FeV] & 3891 &54.8& 0.74  \\
$[$FeVI] & 5176 &75.0& 0.62 &$[$NiII] & 4326 &7.6& 0.35 & [NiIII]   & 6000      &18.2& 0.05 \\
 \hline
\end{tabular}}
\caption{\footnotesize{The forbidden lines suggested to have the Zeeman effect in strong magnetic field. IP is the ionization potential, and $A$ is the Einstein absorption coefficient. The parameters are taken from Allen's astrophysical quantities in the fourth edition.}}
\end{table}

\end{document}